\documentclass{emulateapj} 

\usepackage{natbib} 
\usepackage[figuresright]{rotating} 
\usepackage{graphicx} 
\usepackage{subfigure} 
\usepackage{psfig} 
\newcommand{\hei}{He~{\sc i} 2.112 $\mu m$} 
\newcommand{\heia}{He~{\sc i} 2.058 $\mu m$} 
\newcommand{\brg}{Br$\gamma$} 
\newcommand{\mum}{\ifmmode \mu m \else $\mu m$\fi} 
 
\newcommand{\teff}{\ifmmode T_{\rm eff} \else $T_{\mathrm{eff}}$\fi} 
\newcommand{\logg}{\ifmmode \log g \else $\log g$\fi} 
\newcommand{\lL}{\ifmmode \log \frac{L}{L_{\odot}} \else $\log \frac{L}{L_{\odot}}$\fi}

\newcommand{\msun}{\ifmmode M_{\odot} \else M$_{\odot}$\fi} 
\newcommand{\zsun}{\ifmmode Z_{\odot} \else Z$_{\odot}$\fi} 
\newcommand{\lsun}{\ifmmode L_{\odot} \else L$_{\odot}$\fi} 
\newcommand{\rsun}{\ifmmode R_{\odot} \else R$_{\odot}$\fi} 
 
\slugcomment{}

\shorttitle{Top-Heavy IMF in the Galactic Center Stellar Discs} 
\shortauthors{Bartko et al.}

\begin{document}

\title{An Extremely Top-Heavy IMF in the Galactic Center Stellar Disks 
\footnote{Based on observations collected at the ESO Very Large Telescope (programs 075.B-0547, 076.B-0259 and 077.B-0503)}} 



\author{ 
H.~Bartko\altaffilmark{a,}\altaffilmark{*}, 
F.~Martins\altaffilmark{a,b,*},  
S.~Trippe\altaffilmark{j} 
T.~K.~Fritz\altaffilmark{a},  
R.~Genzel\altaffilmark{a,c},  
T.~Ott\altaffilmark{a},  
F.~Eisenhauer\altaffilmark{a},  
S.~Gillessen\altaffilmark{a},  
T.~Paumard\altaffilmark{e},  
T.~Alexander\altaffilmark{f}, %
K.~Dodds-Eden\altaffilmark{a},  
O.~Gerhard\altaffilmark{a}, %
Y.~Levin\altaffilmark{d}, %
L.~Mascetti\altaffilmark{a},  
S.~Nayakshin\altaffilmark{g}, %
H.~B.~Perets\altaffilmark{f}, %
G.~Perrin\altaffilmark{e}, %
O.~Pfuhl\altaffilmark{a},  
M.~J.~Reid\altaffilmark{h},  
D.~Rouan\altaffilmark{e}, %
M.~Zilka\altaffilmark{i}, 
A.~Sternberg\altaffilmark{i}, 
} 
 
 
 
 \altaffiltext{a} {Max-Planck-Institute for Extraterrestrial Physics, Garching, Germany} 
 \altaffiltext{b} {GRAAL-CNRS, Universit' Montpelier II - UMR5024, Place Eugène Bataillon, F-34095, Montpellier, France} 
 \altaffiltext{c} {Department of Physics, University of California, Berkeley, USA} 
 \altaffiltext{d} {Leiden University, Leiden Observatory and Lorentz Institute, , NL-2300 RA Leiden, the Netherlands} 
 
 \altaffiltext{e} {LESIA, Observatoire de Paris, CNRS, UPMC, Université Paris Direrot, Meudon, France} 
 \altaffiltext{f} {Faculty of Physics, Weizmann Institute of Science, Rehovot 76100, Israel} 
 \altaffiltext{g} {Department of Physics \& Astronomy, University of Leicester, Leicester, UK} 
 \altaffiltext{h} {Harvard-Smithsonian Center for Astrophysics, 60 Garden Street, Cambridge, USA} 
 \altaffiltext{i} {School of Physics and Astronomy, Tel Aviv University, Tel Aviv 69978, Israel} 
 \altaffiltext{j} {IRAM Grenoble, 300 rue de la piscine, F-38406 Saint Martin d'Heres, France} 
 \altaffiltext{*} {correspondence: H.~Bartko, hbartko@mpe.mpg.de; F.~Martins, fabrice.martins@graal.univ-montp2.fr}

\begin{abstract} 
We present new observations of the nuclear star cluster in the central parsec of the Galaxy with 
the adaptive optics assisted, integral field spectrograph SINFONI on the 
ESO/VLT. Our work allows the spectroscopic detection of early and late type stars to $m_K\ge 16$, 
more than 2 magnitudes deeper than our previous data sets. 
Our observations result in a total sample of 177 bona fide early-type stars. 
We find that most of these Wolf Rayet (WR), O- and B- stars reside in two strongly warped disks between 
0.8'' and 12'' from SgrA*, as well as a central compact concentration (the S-star cluster)  
centered on SgrA*. The later type B stars ($m_K>15$) in the radial interval between 0.8'' and 12'' seem to be in a more isotropic distribution outside the disks. 
The observed dearth of late type stars in the central few arcseconds is puzzling, even when allowing for stellar collisions.
The stellar mass function of the disk stars is extremely top heavy with a best fit power law of $\mathrm{d}N/\mathrm{d}m \propto m^{-0.45\pm0.3}$.
Since at least the WR/O-stars were formed in situ in a single star formation event $\sim$6 Myrs ago, this mass function 
probably reflects the initial mass function (IMF). 
The mass functions of the S-stars inside 0.8'' and of the early-type stars at distances beyond 12'' are compatible with a standard Salpeter/Kroupa IMF  
(best fit power law of $\mathrm{d}N/\mathrm{d}m \propto m^{-2.15\pm0.3}$).
\end{abstract} 
 
 

\section{Introduction} 
\label{intro} 
 
The central parsec of the Galaxy harbors more than one hundred young massive stars  
\citep{Forrest1987,Allen1990,Krabbe1991,Najarro1994,Krabbe1995,Blum1995,Tamblyn1996,Najarro1997,Genzel2003,Ghez2003,Paumard2006,Martins2007}. 
This is highly surprising since the tidal forces from the central four million solar mass black hole should make formation of stars by gravitational collapse from a cold interstellar cloud very difficult if not impossible \citep{Morris1993}. 
Observations have established that most of the WR-stars and 
O-stars (dwarfs, giants, and supergiants) dominating the luminosity of the 
early-type population were formed in a well defined single event $\sim$6 Myrs ago, perhaps as the result 
of the infall of a gas cloud followed by an in-situ star formation event  
\citep{Krabbe1995,Paumard2006}. About half of these WR/O-stars  
between 0.8'' and 12'' from SgrA* appear to reside in a well defined but highly warped 
disk that rotates clockwise on the sky \citep{Levin2003,Genzel2003,Paumard2006,Lu2009,Bartko2009}. It is 
less clear how the other half is distributed. \citet{Paumard2006} and \citet{Bartko2009} find evidence 
for a counter-clockwise structure, perhaps a second disk in a dissolving state.  
\citet{Lu2009} confirm the first stellar disk but do not observe a significant number of  
stars in the second disk. Within about 1'' of SgrA* there is a sharp cutoff in the density of 
WR/O-stars. Instead there is a concentration of fainter stars with randomly oriented and eccentric 
orbits: the so-called 'S-star cluster' \citep{Eisenhauer2005,Gillessen2009}.  
The brighter members of this central cusp are spectroscopically identified as main 
sequence B stars \citep{Ghez2003,Eisenhauer2005,Martins2008,Gillessen2009}.  
It is currently debated whether the S-stars formed in the disk(s) and subsequently migrated to the central arcsecond \citep{Levin2007,Loeckmann_etal2008}, or whether they formed outside the central few parsecs, were injected into near-parabolic orbits by massive perturbers and were then captured by the massive black hole \citep[e.g.][]{Hills1988,Perets2007}. 
 
If star formation indeed has taken place in the vicinity of the massive black hole, 
it is of great interest to explore the stellar mass function and spatial distribution to gain a better 
understanding of the processes involved in overcoming the tidal barrier. Likewise it is important to 
observationally constrain the characteristics and origin of the central S-star cluster. 
In the following we present the results of an extensive imaging spectroscopy survey of the 
central parsec with the adaptive optics (AO) assisted integral field spectrograph SINFONI on the ESO/VLT. 
We report new observations aimed at defining the spatial distribution and dynamics of the nuclear stars up to $m_K\geq16$ throughout 
this region and combine this new work with our earlier results on the WR/O-stars and the central S-star cluster.

 
\section{Observations and Data Reduction} 
\label{observ}


\begin{figure}[t!] 
\begin{center} 
\vspace{0.3cm}
\includegraphics[width=\columnwidth]{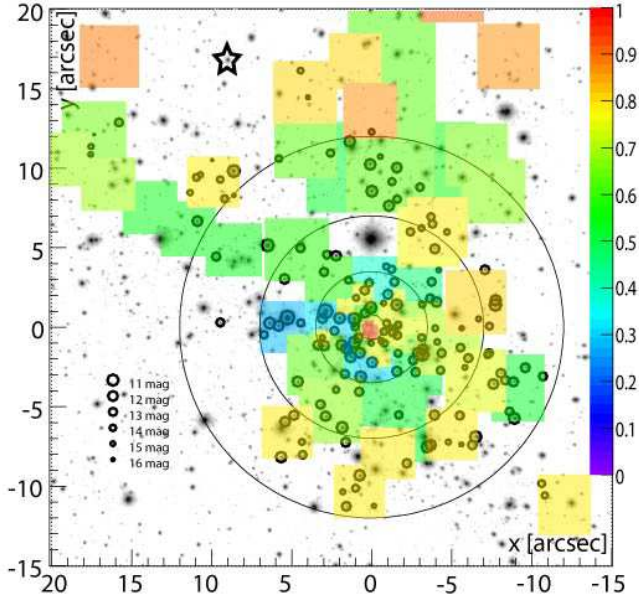}
\caption{\it \small The black circles indicate the sample of early-type stars with $0.8''\leq R \leq 25''$. The size of the circle scales with the K-band magnitude. The large black circles show projected distances of 0.8'', 3.5'', 7'' and 12'' from Sgr~A*. The colored squares indicate the exposed fields with SINFONI in the 25~mas/pixel and 100~mas/pixel scale. The color scale indicates the combined photometric and spectroscopic completeness for an $m_K=14$ main sequence early-type star in the observed field. The whole inner $\sim12''$ region is contained in lower resolution (250~mas/pixel scale) SINFONI observations \citep{Paumard2006}. The asterisk indicates the visible AO guide star.
} 
\label{fig:comp} 
\end{center} 
\end{figure}

Observations of the imaging survey were carried out from 2005 to 2009 with SINFONI \citep{Eisenhauer2003}  
on the ESO/VLT, partly using natural guide star adaptive optics (AO) and partly exploiting the laser guide star  
facility \citep{Rabien2003,Bonaccini2006}.  
We covered a total surface area of 15 square arcseconds at projected distances below 4'' from SgrA*  
with the AO scale ($12.5 \times 25$ mas/pixel scale) resulting in a final K-band full  
width at half maximum (FWHM) of typically about 100~mas. 
In addition, we observed a total surface area of more than 500 square arcseconds with the $50 \times 100$~mas/pixel scale  
resulting in typical K-band FWHMs of about 200~mas. The location of the observed fields (typical integration time per field 2 hours) 
is indicated in Figure \ref{fig:comp}. 
 
We reduced the data with the software SPRED as in \citet{Eisenhauer2005} and \citet{Paumard2006}.  
Our primary goal in this project is to determine the spatial distribution and dynamics of the early type stars as a function of their K-band magnitude / mass. For this purpose we need to distinguish from the spectroscopic data early type stars from late type stars and to isolate the central star cluster from fore- and background stars. 

Late-type stars in the observed magnitude range are giants later than K0III and cool red supergiants. These stars show characteristic CO band head absorption features, which can efficiently be used to remove these stars from our sample.
Only three red supergiants have been observed in the central pc of the Galaxy \citep{Blum2003,Paumard2006}, most notably IRS 7 at an angular distance of 5.5" from SgrA*. In contrast to the giants later than K0III the red supergiants are young and massive stars with ages between 7 and a few tens of Myrs \citep{Blum1996}.

We classified stars as ``CO'' (late type) or ``non CO'' (early-type) depending on the presence or absence of the CO 2.29$\micron$ $v=0-2$ band-heads 
 \citep[see also][]{Maness2007}. 
By adding artificial late type stars to our data we determined the average efficiency of the CO detection to be better than 98\% at $m_K=16.5$.
In order to rate a ``non CO'' as bona fide early-type star we require the detection of at least one of the \brg\ , \heia\ ,or \hei\ lines at a signal to noise level, which allows a radial velocity measurement with a maximum uncertainty of 100 km/s. The sample of bona fide early-type stars is essentially free from late type star interlopers.
 
Fore- and background early-type stars do not add any significant contamination to the observed population of early-type stars in the central parsec around SgrA*. 
Due to the large extinction toward the Galactic Center of $A_K = 2.8$ \citep[see e.g.][]{Schoedel2007}, foreground sources can efficiently be rejected by their very blue H-K band colors. There are only two bright infrared detectable foreground stars \citep{Biretta1982}. Background sources are subject to an even higher
extinction and very red H-K band colors.
The cusp of early-type stars around SgrA* is very steep (see Figure \ref{fig:dens_prof}): $\Sigma(R) \propto R^{-1.4\pm0.2}$. For a cluster in equilibrium the distribution of the line-of-sight distances to SgrA* is related to the star surface number density distribution \citep{Alexander2005}. Out of the 136 early-type stars in the interval of projected distances $0.8''\leq R \leq 12''$ only 8 / 3 stars may have line-of-sight distances larger than 1 / 2 pc.

We corrected the observed K-band magnitude $m_{K,\mathrm{obs}}$ \citep[for the absolute magnitude scale calibration see][]{Blum1996,Maness2007} of the stars  
for the variation of extinction over the Galactic Center region. 
We calculated the extinction for each position in the Galactic Center field, $A_{K,\mathrm{obs}}$, by using the observed H band and K-band magnitudes of the 20 nearest neighboring stars, assuming realistic  
intrinsic colors for the stars and assuming the extinction law from \citet{Rieke1989}, $A(\lambda) \propto \lambda^{-1.75}$. All magnitudes $m_K$ mentioned below are referred back to a common standard extinction of 2.8: $m_K = m_{K,\mathrm{obs}}+2.8-A_{K,\mathrm{obs}}$.

Our total data set contains 177 bona fide early-type stars at projected distances up to 25'' from SgrA*. 
Figure \ref{fig:comp} shows the location of the bona fide early-type stars with $R>0.8''$.
Of the 177 bona fide early-type stars in our survey 
118 are WR/O-stars and 59 are B-dwarfs ($m_K \ge 14.$); 28 WR/O-stars and 34 B-dwarfs are reported here for the first time).

\newpage 
\section{Number Counts and Completeness Corrections} 
\label{completeness}

 
To study the radial surface density and KLF of the early-type stars, we correct the number of observed stars  
for both photometric and spectroscopic incompleteness, as well as (for radial surface density plots) for incomplete areal coverage. 
 
The photometric completeness $\epsilon_{\mathrm{phot}}$ is defined as the probability of detecting a star with a given magnitude in the image generated by integrating over the spectral information of the SINFONI data.  
For this purpose, we added artificial stars to the spectrally integrated SINFONI images, and determined the probability  
of re-detecting the artificial stars with an automatic star detection algorithm \citep{Stetson1987}. 
We used a dense sampling of the image area with a satisfying number of trials (about 2000 trials per image and per flux bin, the exact number depending on respective image dimensions) and tested the completeness was in flux bins of $\Delta m_K = 1$. 
 
The spectroscopic completeness $\epsilon_{\mathrm{spec}}$ is the probability of  identifying the spectral type of a given, photometrically detected star in 
a data cube. For this purpose we added and recovered 3D stellar cubes in analogy with the 
photometric completeness estimates. These 3D cubes are formed from 
a spatial PSF with an early-type star spectrum along the wavelength 
axis. 
For each photometrically re-detected star we extracted its spectrum from the data cube and computed a CO index \citep{Maness2007}, which quantifies the relative depth of the CO band head absorption feature. We required a minimum total signal to noise ratio of the spectrum in the region of the CO band head of at least 5 sigma. 
If the relative depth of the CO band head feature of the artificial early-type star spectrum was smaller than twice the noise, the star was counted as a ``spectroscopically re-identified early-type star". 
 
 
 
We assign errors to both photometric and spectroscopic completeness values. We confirmed that the completeness values do not change beyond these errors by different settings of the star detection algorithm and the CO index calculation. We computed the combined photometric and spectroscopic completeness as a function of magnitude and location throughout the observed region, propagating the individual errors, to the combined completeness $\epsilon_{\mathrm{comb}}(x,y,m_K)$. 
We calculated completeness maps, which give (as a function of magnitude) the
completeness and its error for each observed point.
As an example, Figure \ref{fig:comp} shows the combined completeness distribution for an $m_K=14$ main sequence early-type star.  
In our survey an average 50\% completeness is achieved for an $m_K=15$ star. The spectroscopy is deeper toward the central S-star cluster 
and several fields without bright stars (Figure \ref{fig:comp}). 


We derived an effective exposed area $A_{\mathrm{eff}}(R_1,R_2,m_K)$ (including error) over an
annulus between $R_1$ and $R_2$ around SgrA* as a function of $m_K$ taking the 
SINFONI exposure locations and the completeness values including their errors into account: 

\begin{equation}
A_{\mathrm{eff}}(R_1,R_2,m_K) = \int_{\sqrt{x^2+y^2} \geq R_1}^{\sqrt{x^2+y^2} \leq R_2} \epsilon_{\mathrm{comb}}(x,y,m_K) \mathrm{d}x \mathrm{d}y \ .
\end{equation}

Finally, we computed from the number of observed stars $N_{\mathrm{stars,obs.}}(R_1,R_2,m_K)$ in an annulus between $R_1$ and $R_2$ around SgrA*  the K band luminosity function $\mathrm{KLF}(R_1,R_2,m_K)$ and the surface density $\Sigma(R_1,R_2)$:

\begin{eqnarray*}
\mathrm{KLF}(R_1,R_2,m_K) &=& \frac{ N_{\mathrm{stars,obs.}}(R_1,R_2,m_K)}{A_{\mathrm{eff}}(R_1,R_2,m_K)} \ , \\
%
\Sigma(R_1,R_2) &=& \sum_{m_K}{\frac{ N_{\mathrm{stars,obs.}}(R_1,R_2,m_K)}{A_{\mathrm{eff}}(R_1,R_2,m_K)}} \ .
\end{eqnarray*} 
%



 

\section{Results} 
\label{sec:results}

\begin{figure}[t!] 
\begin{center} 
\vspace{0.3cm}
\includegraphics[totalheight=8cm]{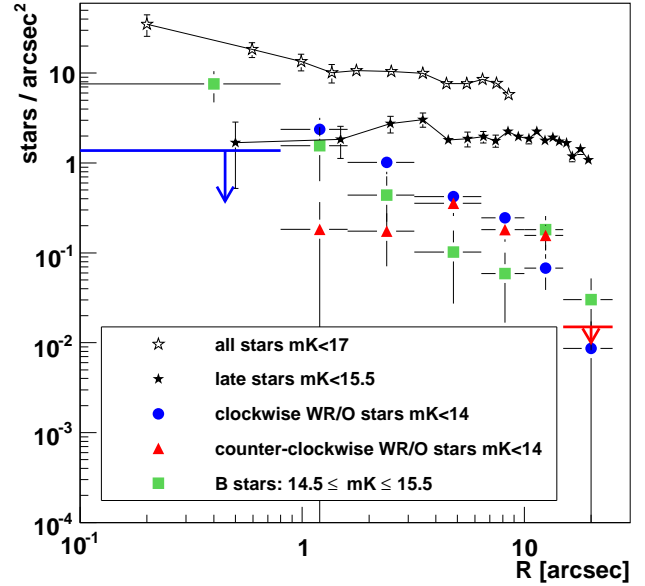} 
\caption{\it \small Projected, completeness, and coverage corrected surface density as a function of distance to SgrA*. Blue points (red triangles) indicate WR/O ($m_K<14$) stars on clockwise (counter-clockwise) orbits and green squares indicate B-dwarfs in the magnitude bin $14.5\leq m_K \leq 15.5$. The blue and red arrows show the 90\% confidence limit to the number density of WR/O-stars in the innermost 0.8'' and counter-clockwise WR/O-stars in the radial interval between 15'' and 25''. For comparison, the open asterisks indicate all stars with $m_K <17$ and the full asterisks the identified late type stars with $m_K<15.5$. 
} 
\label{fig:dens_prof} 
\end{center} 
\end{figure}

 
\subsection{Surface density profiles} 
\label{s_dens_prof}

 

Figure \ref{fig:dens_prof} shows the projected, completeness, and coverage corrected surface density profiles for all  
stars with $m_K<17$, late type stars with $m_K<15.5$, the WR/O-stars ($m_K<14$), and B-dwarfs in the magnitude interval $14.5 \leq m_K \leq 15.5$.  
The radial surface density of the clockwise WR/O-stars follows a power law $\Sigma(R) \propto R^{-1.4\pm0.2}$ between 0.8'' and 15''. 
Extrapolating this power law to smaller projected distances yields $28\pm8$ clockwise WR/O stars for $R<0.8"$. At the 95\% confidence level we expect more than 12 stars within 0.8".
We did not observe any WR/O star in this region. The Poisson probability to observe no star for 28 expected stars is $5.6\cdot10^{-13}$ and the Poisson probability to observe none of $12$ expected stars is still $6\cdot 10^{-6}$. Therefore we conclude that there is a significant cutoff in the distribution of WR/O stars at R=0.8".
The difference compared to the $R^{-2}$ power law suggested by earlier work \citep{Paumard2006,Lu2009,Bartko2009} is due to the thorough completeness and coverage correction applied here.
The surface density distribution of the counter-clockwise WR/O-stars is rather flat between 0.8'' and 15'', mainly because it has a larger central 'hole' than the clockwise stars and a maximum in the distribution at around 4.5''.  
The radial surface density profile of B-dwarfs in the magnitude interval $14.5 \leq m_K \leq 15.5$ drops smoothly 
from the central 0.8'', where the S-stars reside, out to about 25\arcsec.  
Its distribution is similar to the clockwise WR/O-stars with a best fitting power law of $\Sigma(R) \propto R^{-1.5\pm0.2}$. In strong contrast to the early-type stars, late type stars with $m_K \leq 15.5$ exhibit a flat distribution inside of 10'', in excellent agreement with \citet{Buchholz2009} and \citet{Do2009}. 
 

 
\subsection{K Band Luminosity Function} 

\begin{figure}[t!] 
\begin{center} 
\includegraphics[totalheight=8cm]{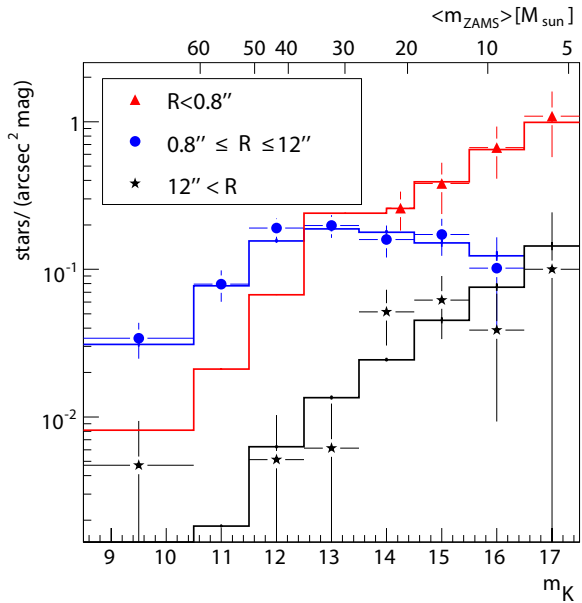} 
\caption{\it \small Completeness corrected K-band luminosity functions of early-type stars in three radial  
intervals $R<0.8''$ (red triangles, scaled by a factor 0.05), $0.8'' \leq R \leq 12''$ (blue points) and 
 $12''<R<25''$ (black asterisks). For a 6 Myr population the best fitting IMF in the radial interval $0.8'' \leq R \leq 12''$, where  
the disks of early-type stars are most prominent, is extremely top heavy and clearly different from the 
 IMFs of the S-stars and the field stars beyond 12''. It can be fitted by a power-law IMF with of $\mathrm{d}N/\mathrm{d}m \propto m^{-0.45\pm0.3}$ or the IMF proposed by \citet{Bonnell2008}.  
The IMF of the field stars beyond 12'' as well as the S-stars within 0.8'' can be fitted by a power-law IMF of $\mathrm{d}N/\mathrm{d}m \propto m^{-2.15\pm0.3}$, consistent with a standard Salpeter/Kroupa IMF. These latter KLFs can also be fitted by somewhat older 
populations with  a continuous star formation history, which would then, however, predict lower numbers of $m_K\leq14$ stars (relevant for S-star cluster). 
The top x-axis shows the average initial
mass of the early-type stars contributing to the K band magnitude bins for the
case of a starburst 6 Myrs ago with an IMF with a slope of $\mathrm{d}N/\mathrm{d}m \propto m^{-0.45\pm0.3}$ as generated with the STARS code. 
Most of the stars with $m_K>14$ are B dwarfs, the brighter magnitude bins contain O dwarfs, giants and WR stars. The main sequence ends at about $m_K=12$, 
corresponding to an initial stellar mass of about $25 M_{\odot}$.
} 
\label{fig:IMF} 
\end{center} 
\end{figure}


Figure \ref{fig:IMF} shows the completeness corrected K-band luminosity functions of early-type stars  
in three radial intervals $R<0.8''$, $0.8'' \leq R \leq 12''$ and $12''<R<25''$. 
The slope of the KLF is sensitive to the stellar mass function (MF). If the population formed in situ without being affected by dynamical effects that might alter the mix of stellar masses in the observed sample, the observed KLF of a young population is directly related to the initial stellar mass function (IMF). 

The stellar disks (best defined in the radial interval $0.8'' \leq R \leq 12''$, see Figure 2) and the randomly orbiting S-stars (in the radial interval $R<0.8''$) are dynamically distinct. The S-stars have a different formation process than the stellar disks \citep{Perets2009}. 

The two main scenarios discussed in the literature for the origin of the disk-like structurse of young stars in the GC are the {\it infalling cluster} model  \citep{Gerhard2001,McMillan2003,Portegiesetal2003,Kim2003,Kim2004,Guerkan2005} and the {\it in situ} star formation model \citep{Levin2003,Genzel2003,Goodman2003,Milosljevic2004,Nayakshin2005,Paumard2006,Bonnell2008,Mapelli2008,Hobbs2009}.

In the {\it infalling cluster} scenario, a cluster of young stars formed at a distance of at least a few parsecs from the GC and then spiraled inward by dyanmical friction. As it finally enters the central parsec it is tidally disrupted and forms a disk-like configuration. If the cluster is dense enough to undergo significant internal mass segregation during the in-spiral phase this scenario does predict a fairly flat MF for the innermost deposited stars even if the assumed intrinsic IMF of the cluster is a standard one \citep[MF: $\mathrm{d}N/\mathrm{d}m \propto m^{-1.1}$,][]{Guerkan2005}. This is because mass segregation within the cluster leads to a differential shedding of stars of different masses as the cluster is tidally stripped during in-spiral. Lower mass stars are deposited further out, while the most massive stars sink the farthest in, especially if the cluster contains a central intermediate mass black hole \citep{Guerkan2005}. The differential shedding of different stellar masses combined with the predicted flat surface density of the deposited cluster stars \citep[$\Sigma \propto R^{-0.75}$,][]{Berukoff2006} would then lead to a strong radial change in the MF and a large ``sea'' of B-stars outside the region where most of the O/WR-stars are located in the plane of the stellar disks.
No such gradient is observed in the main part of the disk ($R<12"$). Figure 3 does show that the KLF and MF of the stars outside 12" steepens, consistent with a normal IMF.
The observed B-stars at $R>12''$, however, have large angular offsets to the disk angular momentum directions and are compatible with isotropic orbits, indicating that these stars were not formed in the same way as the disk stars.
In addition, the steep surface density distribution excludes that there are enough B-stars at large distances to make up for the shortage of such stars in the central 12": $49.3\pm5.7$ stars in the magnitude range $13.5 \leq m_K \leq 16.5$ would be needed (assuming a standard Salpeter/Kroupa IMF) in the radial interval $12'' \leq R \leq 25''$, but only 13 stars have been observed.


In the {\it in-situ model}, a clump or clumps of gas fall into the GC, where they form a disk-like structure. The gaseous disk then fragments to form stars. The detailed observations by \citet{Paumard2006,Bartko2009,Lu2009} are compatible with an {\it in situ} formation of these WR/O stars in a clockwise rotating disk and another highly inclined counter-clockwise structure, possibly a disk. 
The mass segregation time scales in the GC are much larger than 6 Myr \citep{Alexander2005}.
We therefore assume that the stars generated in-situ in the starburst 6 Myr ago are almost fully contained in the observed sample of early-type stars in the radial interval $0.8'' \leq R \leq 12''$.




Our observations demonstrate clearly that the (I)MF in  
the radial interval $0.8'' \leq R \leq 12''$, where the disks of early-type stars are most prominent (see Figure \ref{fig:dens_prof}), must be extremely top heavy. This MF is also clearly different from the MFs of the S star cluster and from the region outside 12''.  
 
The histograms in Figure \ref{fig:IMF} show three theoretical model luminosity functions with different IMFs, \citep[see e.g.][]{Muench2000,Muench2002}. We computed these models using the population synthesis code STARS \citep{Sternberg1998,Sternberg2003} assuming solar metallicity Geneva tracks, for a cluster age of 6 Myr, and an exponentially decaying star burst with an $1/e$ time-scale of 1 Myr. This is the best fitting age and duration of a single star formation event derived from the Hertzsprung-Russell diagram distribution of the O/B-stars and the ratios of various sub-types of WR-stars \citep{Paumard2006}. 
STARS computes K magnitudes using empirical Schmidt-Kaler bolometric corrections and V-K colors for dwarfs, giants, and supergiants. The predicted fractions of O/B 
main sequence stars and giants as well as WR stars agree well with the observations \citep{Paumard2006}: Stars with $m_K<12$ are WR stars and stars with $m_K>15$ are B main sequence stars. Stars with $m_K\sim13$ are WR stars (66\%), evolved early Bstars (31\%) and late Ostars (3\%). Stars with $m_K\sim14$ are Ostars (15\%) and evolved early Bstars (85\%).
\citet{Martins2007} presented detailed stellar atmosphere modelings of a significant fraction of the post-main sequence blue supergiants and WR stars, which have ages of $4-8$~Myrs and ZAMS masses of at least $60 M_{\odot}$, see also \citet{Ott1999,Martins2006}. The average properties of the generated stars are compatible with the observations.

The KLF in the radial interval $0.8'' \leq R \leq 12''$ can be fitted by a power-law IMF of $\mathrm{d}N/\mathrm{d}m \propto m^{-0.45\pm0.3}$ ($\chi^2=4.6$ for 6 degrees of freedom).  
For comparison, the standard Salpeter/Kroupa IMF has 
a high-mass power law of $\mathrm{d}N/\mathrm{d}m \propto m^{-2.3}$. The IMF found by \citet{Bonnell2008} for their $10^5 M_{\odot}$ cloud simulation also gives a  
reasonable fit to the data ($\chi^2 = 7.6$, 6 degrees of freedom).  
We have also selected only those early-type stars, which have an angular offset of less than 10 degrees from the local disk angular momentum direction \citep[see][]{Bartko2009} and subsequently fitted by the theoretical KLF as a function of IMF slope. The obtained IMF slope agrees with the IMF slope of all stars in the radial interval $0.8'' \leq R \leq 12''$ within errors.

The IMF of the field early-type stars beyond 12'' and the S-stars within 0.8'' can be fitted by a  
power-law IMF with a slope of $\mathrm{d}N/\mathrm{d}m \propto m^{-2.15\pm0.3}$, compatible with a standard Salpeter/Kroupa IMF. These latter KLFs can also be 
fitted by Salpeter/Kroupa IMFs and continuous star formation histories with moderate ages ($\leq 60$ Myrs).

It is straightforward to compute a one-to-one relation between observed K band magnitude and initial mass for stars on the main sequence. However, for evolved stars a range of masses contribute at a given K magnitude. Therefore we do not estimate the initial mass for each star but rather compare the observed distribution of K band magnitudes to the simulated magnitude distribution for different IMFs.
The top x-axis of Figure \ref{fig:IMF} shows the average initial mass of the early-type stars contributing to the K band magnitude bins for a 6 Myr old cluster for an IMF of $\mathrm{d}N/\mathrm{d}m \propto m^{-0.45}$ as generated with the STARS code. 
Most of the stars with $m_K>14$ are B dwarfs, the brighter magnitude bins contain O dwarfs, giants and WR stars. The main sequence ends at about $m_K=12$, 
corresponding  to stars with initial masses of about $25 M_{\odot}$ with main-sequence lifetimes of 6 Myr.

An $m_K=16.5$ early-type star for a 6 Myr old population corresponds approximately to a B5V main sequence star with a ZAMS mass of about $7 M_{\odot}$ (see Figure \ref{fig:IMF}). The most massive stars with an individual mass estimates have ZAMS masses of at least $60 M_{\odot}$ \citep{Martins2006,Martins2007}. 
Our estimated IMF slope of the stellar disks ($\mathrm{d}N/\mathrm{d}m \propto m^{-0.45\pm0.3}$) is therefore at least valid over the mass interval $7-60 M_{\odot}$.

 

\subsection{Orbital Angular Momentum Directions} 
\label{dynamics}

\begin{figure}[t!] 
\begin{center} 
\includegraphics[totalheight=7cm]{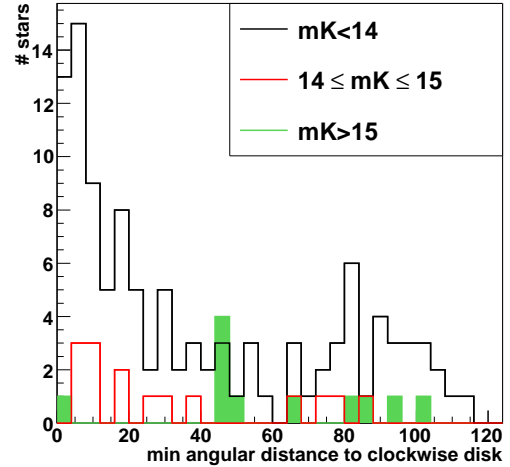}
\caption{\it \small Distribution of the reconstructed angular differences from the local average angular momentum direction of the clockwise system for the 
early-type stars with projected distances between 0.8'' and 12''. The three distributions correspond to different K-band luminosities: black: $m_K<14$, red $14 \leq m_K \leq 15$ and green $mK>15$. Only one out of 11 stars fainter than $m_K=15$ is compatible with the clockwise disk. 
} 
\label{fig:angle} 
\end{center} 
\end{figure}

\begin{figure}[t!] 
\begin{center} 
\includegraphics[totalheight=7cm]{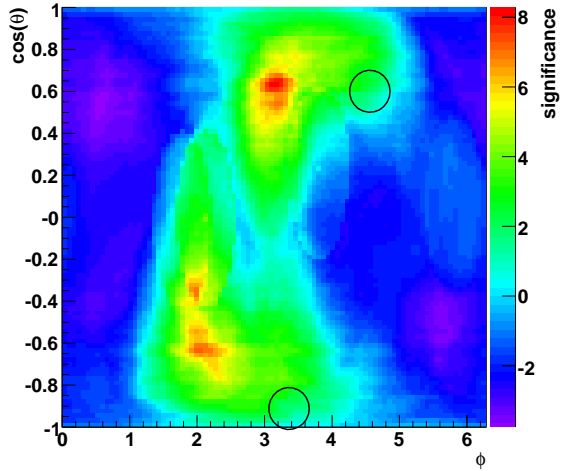} 
\caption{\it \small Cylindrical equal area projection of the distribution of significance in the sky ($25^{\circ}$ aperture, see Bartko et al. 2008) for 82 bona fide early-type stars stars ($m_K<14$) with projected distances between 3.5'' and 15''. The disk positions of Paumard et al. (2006) are marked with full black circles. There are two extended excesses visible for clockwise and counter-clockwise stars, whith maximum significances of 8.2 and 7.1 sigma, respectively. We attribute these two extended excess systems as two warped systems of stars. 
} 
\label{fig:hsky} 
\end{center} 
\end{figure}

Figure \ref{fig:angle} shows the distributions of the reconstructed angular differences from the local average angular momentum direction of the clockwise system, as defined in \citet{Bartko2009}, for the early-type stars with projected distances between 0.8'' and 12''. The three distributions correspond to different K-band luminosities: $m_K<14$, $14 \leq m_K \leq 15$, and $m_K>15$. The peak at small angles corresponds to the clockwise system: 50 out of 110 stars with $m_K<14$ have offsets below $20^{\circ}$ and 8 out of 15 stars with  $14 \leq m_K \leq 15$ have offsets below $20^{\circ}$. In contrast, only one out of 11 stars fainter than $m_K=15$ is compatible with the clockwise disk. This may be an indication that most of the few later B-dwarfs observed in the region of the disks do not belong to the clockwise system but rather to the background population seen at projected distances beyond 12'', which has a more standard IMF. 
 
According to Figure \ref{fig:dens_prof} the clockwise system of massive early-type stars is best defined in the radial range between 0.8'' and 10'' and has a steep surface density distribution. However, most of the counter-clockwise stars reside in the range between 3.5'' and 15'', where we have added considerable SINFONI sky coverage compared to the study of \citet{Bartko2009}. In order to test for the existence of a significant system of counter-clockwise stars, we have computed a sky significance map of angular momentum overdensities for all bona fide early-type stars stars ($m_K<14$) with projected distances between 3.5'' and 15'', see Figure \ref{fig:hsky}. For the technical details see \citet{Bartko2009}. 
There are two extended excesses visible for clockwise and counter-clockwise stars, with maximum significances of 8.2 and 7.1 sigma, respectively. These maximum excesses are clearly offset from the positions found at small projected distances \citep{Paumard2006,Lu2009,Bartko2009} indicating that the two extended excess systems are most probably two warped systems of stars. We confirm a significant counter-clockwise system of stars. 
 

\section{Discussion} 
\label{disc} 
 
The observations discussed in the last section, in combination with the results presented in recent published work 
\citep{Genzel2003,Eisenhauer2005,NayakshinSunyaev2005,Schoedel2007,Paumard2006,Lu2009,Bartko2009,Gillessen2009,Buchholz2009,Do2009} yield the following 
complex and rather unexpected properties of the stellar cluster surrounding the central black hole, 
\begin{itemize} 
\item{most of the early-type stars in the central parsec reside in a combination of a central concentration of main-sequence 
B-stars centered on the black hole (the 'S-star cluster'), plus two strongly warped and almost orthogonal disks (or planar sets of streamers) of WR-, O- and early B-stars. One disk is more  
massive than the second ($\sim10^{4}$ and $\sim5\cdot10^{3}$ $M_{\odot}$, computed from the completeness and exposure corrected number counts and the best fit IMF). The second disk is less well defined, perhaps because it is more disrupted.  
The surface density distribution of the B-stars appears to decline smoothly 
with approximately constant slope (1.5) from the central S-star cluster to the region of the disks. However, both the dynamics 
and the mass function change abruptly at ~1''. Inside this radius, the stellar orbit distribution is random and the ellipticity 
distribution is close to but somewhat hotter than a thermal distribution \citep{Gillessen2009}.  
The observed KLF can be well fitted by a 'standard' 
Salpeter/Kroupa IMF for a single age or continuously star forming population of age a few to 60 Myrs.  
Just outside of this region most of the WR/O-stars 
are in the dominant clockwise disk, which has a 10 degree opening angle and stellar orbits with modest ($e \sim 0.35$) eccentricities. 
The IMF is extremely flat (power law slope 0.45) with a mean stellar mass of 30 $M_{\odot}$. Outside of the region dominated by the two 
disks, at R$\ge 12$'', the KLF appears to again approach that of a 'standard' IMF. Note, however, that even between 1 and 12'' 
$\sim 20-30\%$ of the WR/O-stars, and probably a larger fraction of the B-stars do not appear to belong to either disk system. } 
\item{the late type stars of all luminosities, from the brightest thermally pulsing AGB stars to the faintest, low mass red clump stars, 
do not show a central concentration that might resemble a classical equilibrium stellar 'cusp' as predicted by all theoretical 
studies \citep{Bahcall1976,Bahcall1977,Freitag2006,HopmanAlexander2006}. 
The brightest late type stars show a central hole of radius a few to 7'' \citep{Sellgren1990,Genzel1996,Haller1996} and even 
the faint early K-giants with $m_K\sim15$ show a flat core of radius 10''\citep{Buchholz2009,Do2009}, see Figure \ref{fig:dens_prof}.
The combination of the S-star cluster (mostly unrelaxed early-type stars) and the flat late type star distribution 'conspire' to give a
shallow overall cusp (surface density slope $0.2 ... 0.4$) found in the faint star counts in Figure \ref{fig:comp} 
\citep{Genzel2003,Schoedel2007}}.

The observations suggest that the two WR/O/B-star disks between 0.8'' and 12'' are structurally and dynamically distinct from both
the S-star cluster and the outer region. The properties of these warped disks (or sets of streamers), 
including in particular their steep surface density distribution,
their top heavy IMF, and their relatively low total stellar masses all strongly favor an in situ star formation model \citep{Paumard2006, Lu2009, Bartko2009,Berukoff2006} and broadly agree with the findings in recent hydrodynamical simulations \citep{Bonnell2008,Hobbs2009} of
star formation triggered by gas cloud infall. The sharp transition at 0.8'' between disk zone and S-star cluster, both in terms of dynamics
and mass function, in our opinion also strongly disfavors migration scenarios from the disks for the origin of the S-star cluster.
In turn this sharp transition supports the 'injection and capture' scenarios mentioned in the Introduction. 

The large central core \citep[and perhaps even
central hole, see ][]{Buchholz2009,Do2009} in the density distribution of late type stars of all luminosities is puzzling and currently not understood. There are a number of possible interpretations, none of which at present offers a compelling explanation. As discussed 
by \citet{Do2009} and \citet{Buchholz2009} equilibrium mass segregation by itself cannot account for this distribution \citep{Freitag2006,HopmanAlexander2006}, as even in a multi-mass cluster the lightest stars attain a radial density distribution that is steeper than $R^{-1}$ \citep[probably excluded by the data:][]{Do2009}.
Simulations indicate that physical collisions with main-sequence stars and stellar black holes can remove moderately bright giants in 
the central arcsecond, but not over a much larger region, nor to $m_K \sim 15$ \citep{Dale2009}. Likewise, 
tidal disruption of stars may play a role near the massive black hole \citep{Perets2009} but it needs to be investigated whether they are frequent enough to remove the entire old cusp.
The in-spiral of an intermediate mass
black hole can plausibly gouge out a large enough core in the stellar distribution \citep{MilosavljevicMerritt2001,Baumgardt2006} but there is no
evidence for such a second black hole from the dynamical properties of the S-stars or late type stars, nor from
the motion of SgrA* itself \citep{Gillessen2009,GualandrisMerritt2009,Trippe2008,Reid2004}. 
\citet{Merritt2009} has pointed out that the two-body relaxation time scale probably is significantly longer than the age of the Galactic Center star cluster (and the Hubble time) throughout the central parsec, especially if the spatial distribution of the giants is similar to that of most of the stellar mass. As a result the observed large core may reflect the initial conditions of the Galactic Center nuclear cluster. \citet{Trippe2008} find, however, that the dynamics of the old star cluster is consistent with a relaxed, fully phase mixed system. In addition \citet{Alexander2007} has pointed out the importance of relaxation processes much faster than the standard two-body rate \citep[including the massive perturbers proposed by][]{Perets2007}. The top-heavy IMF discussed in this paper (if applicable to early star formation episodes) would obviously also lead to a lack of old low mass giants in the core. To explain the flat or even inverted radial slope \citep[][ this paper]{Do2009,Buchholz2009}, the IMF probably would have to depend strongly on radius.
The remarkable properties of the Galactic Center nuclear
star cluster remain puzzling and continue to give us food for thought and further work.
 
\end{itemize}

\acknowledgments

We thank all the ESO staff for their help during the various observing runs, and Melvyn Davies for a discussion on his work on stellar collisions. 
We thank the DFG for support via German-Israeli Project Cooperation grant STE1869/1-1.GE625/15-1.
We thank the referee, Don Figer, for his comments and suggestions.
 


\end{document}